\input{aipcheck}

\documentclass[
    ,final            
  ]
  {aipproc}

\layoutstyle{6x9}

\begin{document}

\title{Coronal activity \newline with XMM-Newton and Chandra}


\author{B. Stelzer}{
  address={INAF - Osservatorio Astronomico di Palermo, Piazza del Parlamento 1, I-90134 Palermo, Italy}
}

\begin{abstract}
{\em XMM-Newton} and {\em Chandra} have greatly deepened our knowledge of stellar coronae 
giving access to a variety of new diagnostics 
such that nowadays a review of stellar X-ray astronomy necessarily must focus on a few 
selected topics. Attempting to provide a limited but representative overview of recent discoveries 
I discuss three subjects: the solar-stellar connection,
the nature of coronae in limiting regimes of stellar dynamos, and "hot topics" on 
X-ray emission from pre-main sequence stars.  
\end{abstract}

\maketitle


\section{The solar-stellar analogy}

Ever since the detection of X-ray emission from "normal" stars a comparison of their X-ray properties
to those of the Sun has been at the heart of stellar X-ray astronomy. 
Some of the recent discoveries in investigations of the solar-stellar analogy are 
the first detection of X-ray activity cycles on stars other than the Sun, 
studies of the hard X-ray emission during solar and stellar flares, and  
statistical analysis of flare frequencies.

\subsection{X-ray activity cycles}

Stellar activity cycles have been known to exist from more than 40yrs of monitoring of chromospheric Ca\,II
emission during the Mt.Wilson program \cite{Baliunas95.2}. 
Having in mind that for the Sun, 
the amplitude of the Ca\,II variations throughout its $11$\,yr magnetic cycle is only a factor of 
two, while the X-ray luminosity of nearby stars varies by a factor of $100$, it is reasonable to
suspect the existence of coronal activity cycles on solar-like stars. 
This contrasts with the observation that stars generally show little evidence of long-term X-ray variability. 
Ca\,II activity cycles are generally found on relatively inactive stars, while most stars are
much more active than the Sun and this might explain their different long-term behavior. Based on these
considerations, inactive stars are considered to be the most promising candidates for the detection of
X-ray activity cycles. Dedicated {\em XMM-Newton} monitoring campaigns with snapshot observations on intervals
of $6$\,months have lately 
provided evidence for systematic long-term variability of the X-ray luminosity in
a few nearby stars. 

\begin{figure}
  \parbox{\textwidth}{
  \parbox{0.5\textwidth}{
  \includegraphics[width=0.5\textwidth]{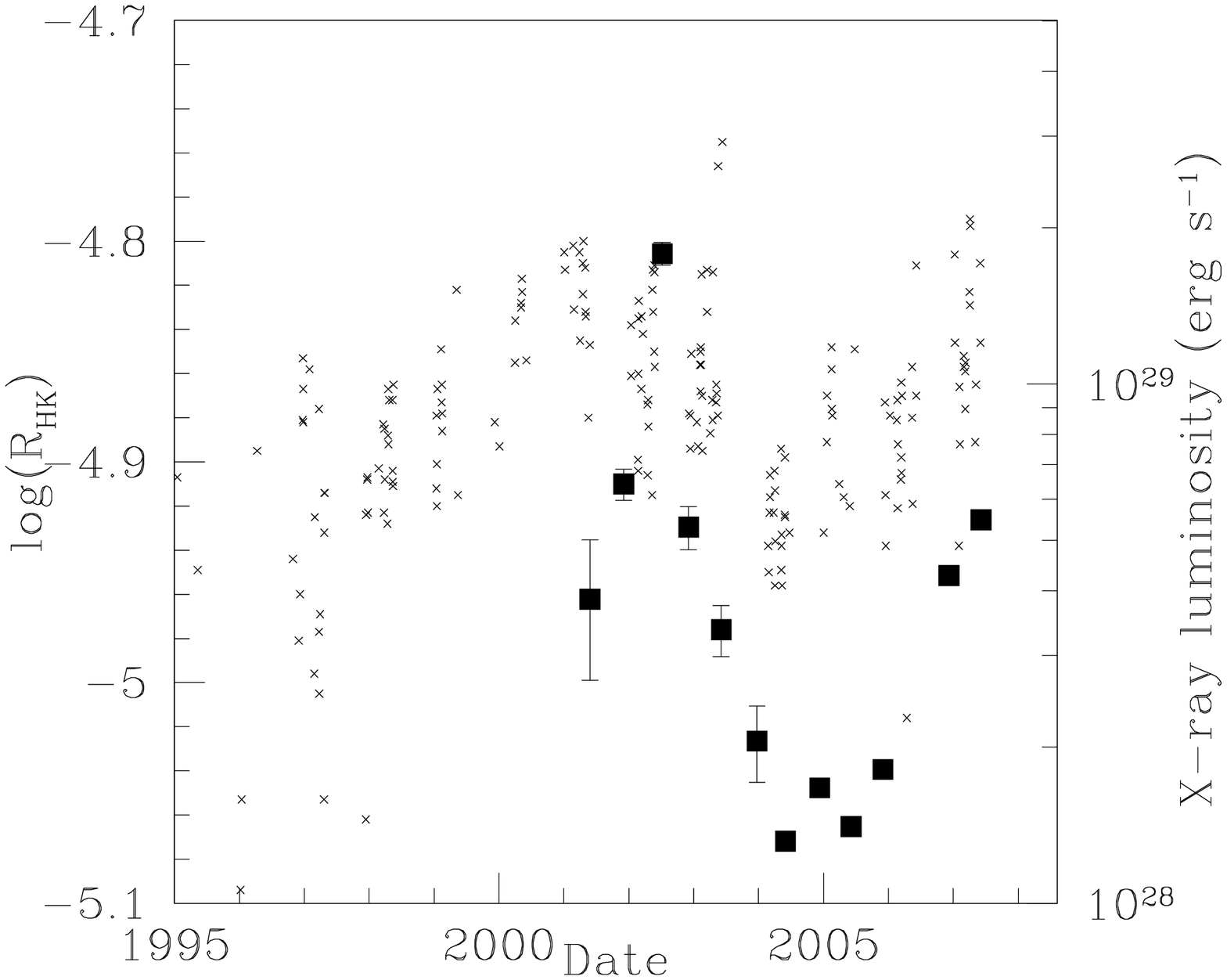}
  }
  \parbox{0.5\textwidth}{
  \includegraphics[width=0.5\textwidth]{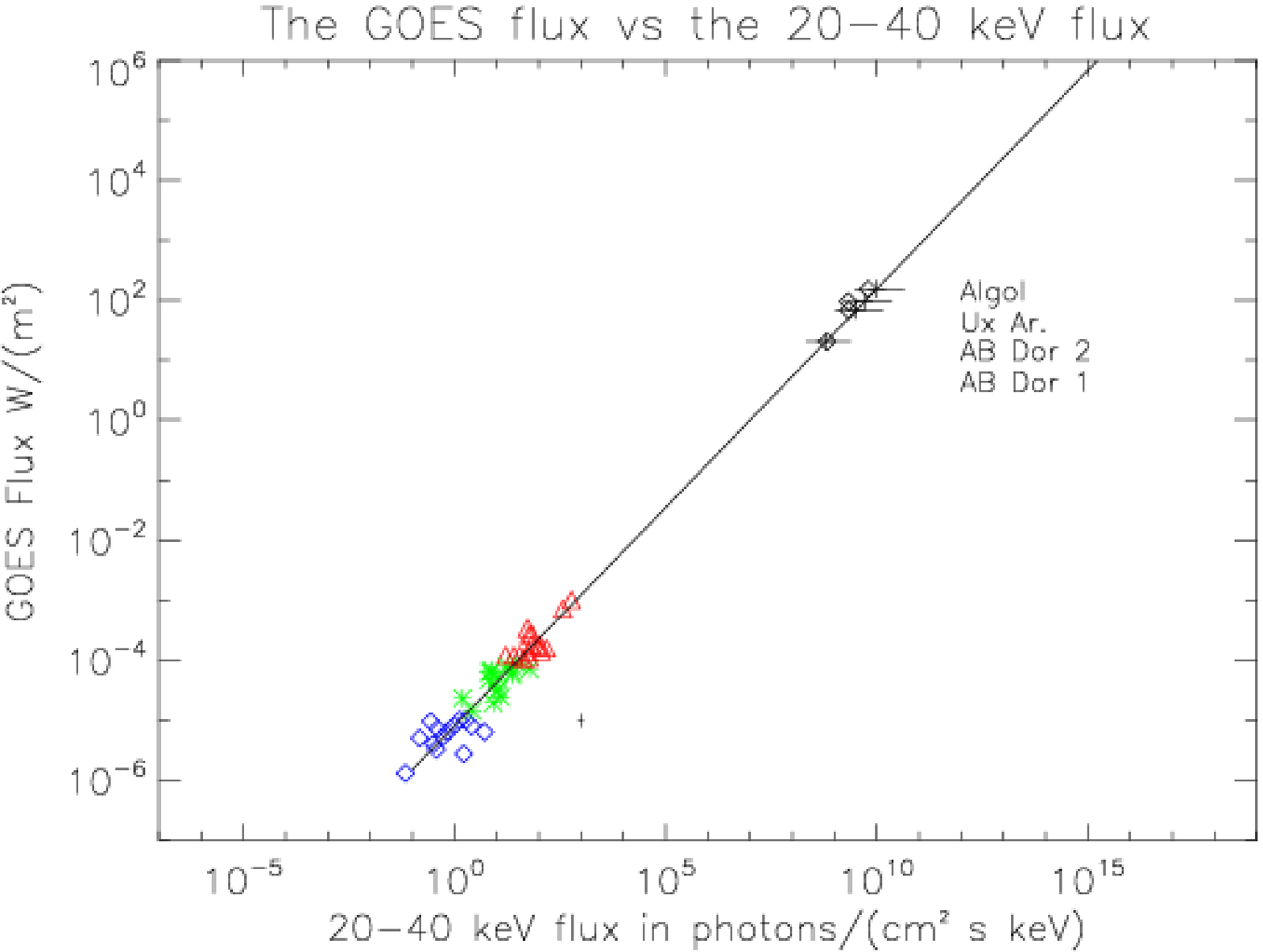}
  }
  }
  \caption{{\em left} - Ca\,II (small crosses) and X-ray (large circles) activity cycles of the low-activity G star HD\,81809 \protect\cite{Favata08.1}; 
{\em right} - Peak flare fluxes at soft and hard X-ray energies for solar and stellar flares \cite{Isola07.1}.}
  \label{fig:1}
\end{figure}

The G-type star HD\,81809, to date observed with {\em XMM-Newton} for a full period of its $8$\,yr Ca\,II cycle, 
displays X-ray variations in accordance with the Ca period (see Fig.~\ref{fig:1}). 
Similar to the case of the Sun, the amplitude in X-ray luminosity exceeds that of the Ca\,II emission by far. 
The X-ray luminosity and temperature in different phases of the cycle extend the trend observed in
solar data, and can be explained by varying surface coverage with cores of active regions \cite{Favata08.1}. 
Direct evidence for X-ray activity cycles from long-term monitoring has been presented for another two stars,
$\alpha$\,Cen and $61$\,Cyg \cite{Robrade05.1, Hempelmann06.1}. 
Systematic variability is observed but not enough data has been
accumulated to establish a periodic pattern so far. 

From the theoretical side, X-ray variability from active stars is predicted 
on basis of the structure of the coronal magnetic field inferred from an extrapolation of the 
surface magnetic field maps for the test case AB\,Dor \cite{McIvor06.2}. 
The X-ray signal produced by different types of underlying star spot distributions was computed 
by synthesizing the X-ray emission in closed regions of the estimated coronal field. 
These simulations show that, depending on the star spot pattern, cyclic variations of the
X-ray emission measure may or may not be present. Clearly, from the absence of an X-ray cycle one can not 
conclude that there is no magnetic cycle.

\subsection{Soft and hard X-rays in flares}

From the standard "thick-target" model for solar/stellar flares one expects a 
direct causal connection between hard (non-thermal) and soft (thermal) X-ray emission \cite{Neupert68.1}. 
The relation between the soft and hard peak fluxes of solar flares was studied by \cite{Isola07.1} using
GOES data for the soft emission and RHESSI data for the hard emission. A power-law relation 
between the peak flux in soft and in hard X-rays was found that holds for over $3$ dex of soft X-ray flux
from solar flare class C to class X. 
The {\it expected} flux relation in the soft and hard emission for a thermal spectrum 
implies a $\sim 6$\,keV plasma, much hotter than what is observed during
solar flares. Therefore, the hard emission was attributed to a non-thermal origin.

Due to the poor sensitivity for hard X-rays of most stellar X-ray instruments, 
very few stellar flares have been observed at energies above $10$\,keV. 
\cite{Isola07.1} evaluated the peak fluxes of a handful of stellar flares observed with {\em BeppoSAX} 
in the same bands that had been defined for the solar flare observations. Fig.~\ref{fig:1} shows 
that the stellar flares line up almost perfectly along the extrapolation of the Sun's power-law relation between
soft and hard peak flare flux. The existence of a unique scaling law for solar and stellar flares 
underlines that flares are a universal phenomenon.

\subsection{Flare number energy distributions}

The origin of the quiescent
X-ray emission of the Sun and the stars has always
been a mystery. In absence of a theoretical explanation for the production of persistent X-rays, is has
been conjectured that what we observe as apparently quiet might actually be a super-position 
of small unresolved flares, termed nano-flares because of their $\sim 10^9$ times lower 
energy compared to the largest solar flares with $10^{33}$\,erg/s \cite{Parker88.1}. 
If so, flares are of fundamental importance for the solar coronal heating problem
and they may be the actual and the only driver of magnetic activity. 

Whether nano-flares 
are sufficient to come up for the thermal energy budget in the corona of the Sun and the stars depends on 
their frequency. The number distribution of flares as a function of energy can be approximated by a power-law,
$\frac{dN}{dE} \sim E^{-\alpha}$. 
For a power-law index $\alpha > 2$ the integrated energy of all flares diverges at the low end, 
and this provides (naively speaking) an infinite heating resource for the corona. 
For the Sun, \cite{Aschwanden00.1} observed a frequency distribution of flare energies 
with a unique slope over eight orders of magnitude in energy \cite[e.g.][]{Aschwanden00.1}.
However, these results are still disputed and values 
between $\sim 1.4...2.6$ have been cited \cite{Berghmans98.1, Krucker98.1}. 

For stars other than the Sun in general there is no sufficient database for statistical flare studies.
A remedy is to look at a sample of stars that can be assumed to be similar, 
e.g. in an open cluster or in a star forming region, 
and analyse their collective flare frequency. Recent comprehensive
X-ray surveys in star forming regions have been exploited for this purpose. 
In particular, flare number energy distributions have been evaluated
for Orion and Taurus making use of two deep X-ray exposures,  
the {\em Chandra Orion Ultradeep Project} (COUP) and the 
{\em XMM-Newton Extended Survey in the Taurus Molecular Clouds} (XEST), 
respectively; see \cite{Getman05.1, Guedel07.3} for details on these surveys. 

\begin{figure}
  \parbox{\textwidth}{
    \parbox{0.4\textwidth}{
      \includegraphics[width=0.4\textwidth]{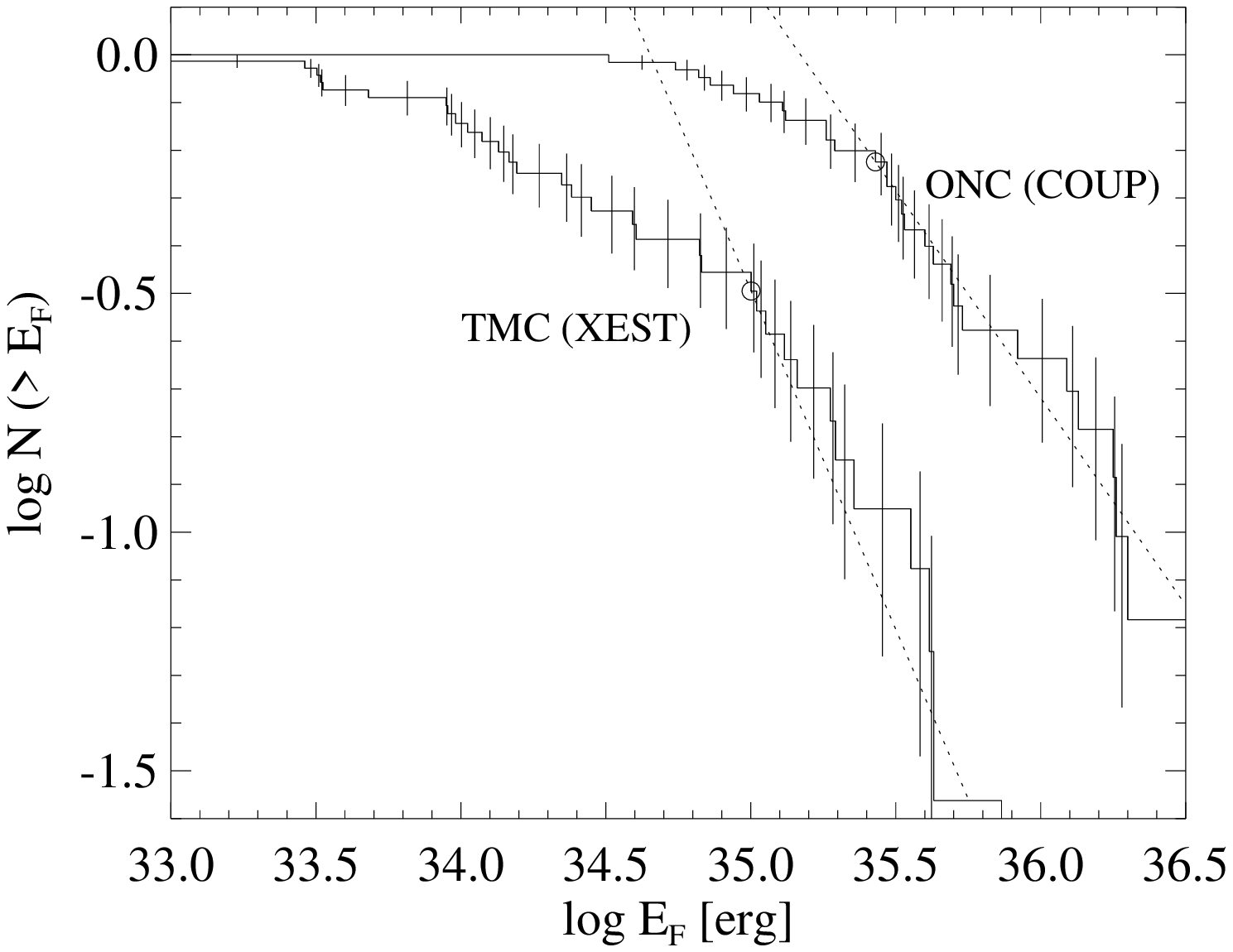}
    }
    \parbox{0.6\textwidth}{
       \parbox{0.29\textwidth}{
         \includegraphics[width=0.29\textwidth,bb=60 160 540 640,clip]{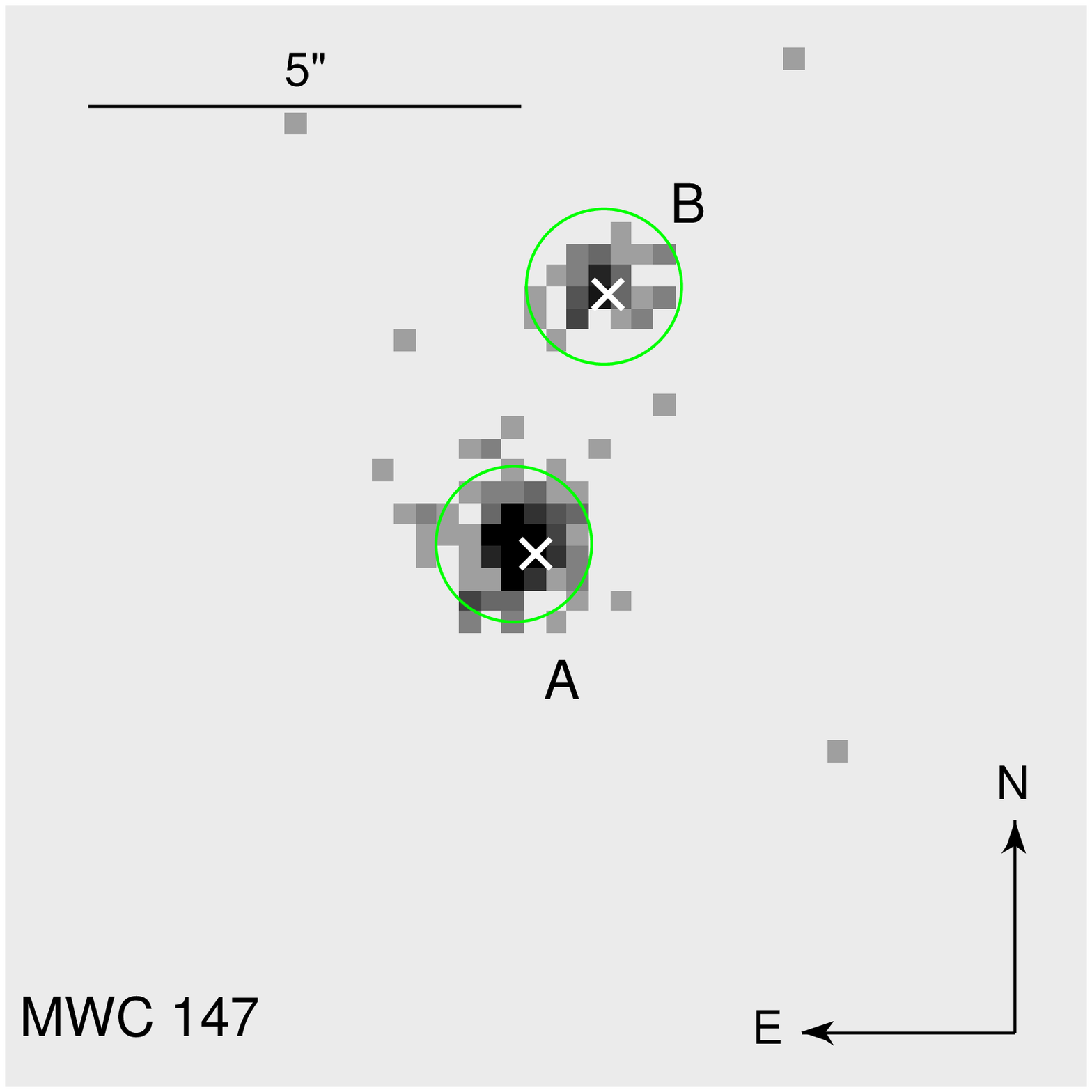}
       }
       \parbox{0.29\textwidth}{
         \includegraphics[width=0.29\textwidth,bb=60 160 540 640,clip]{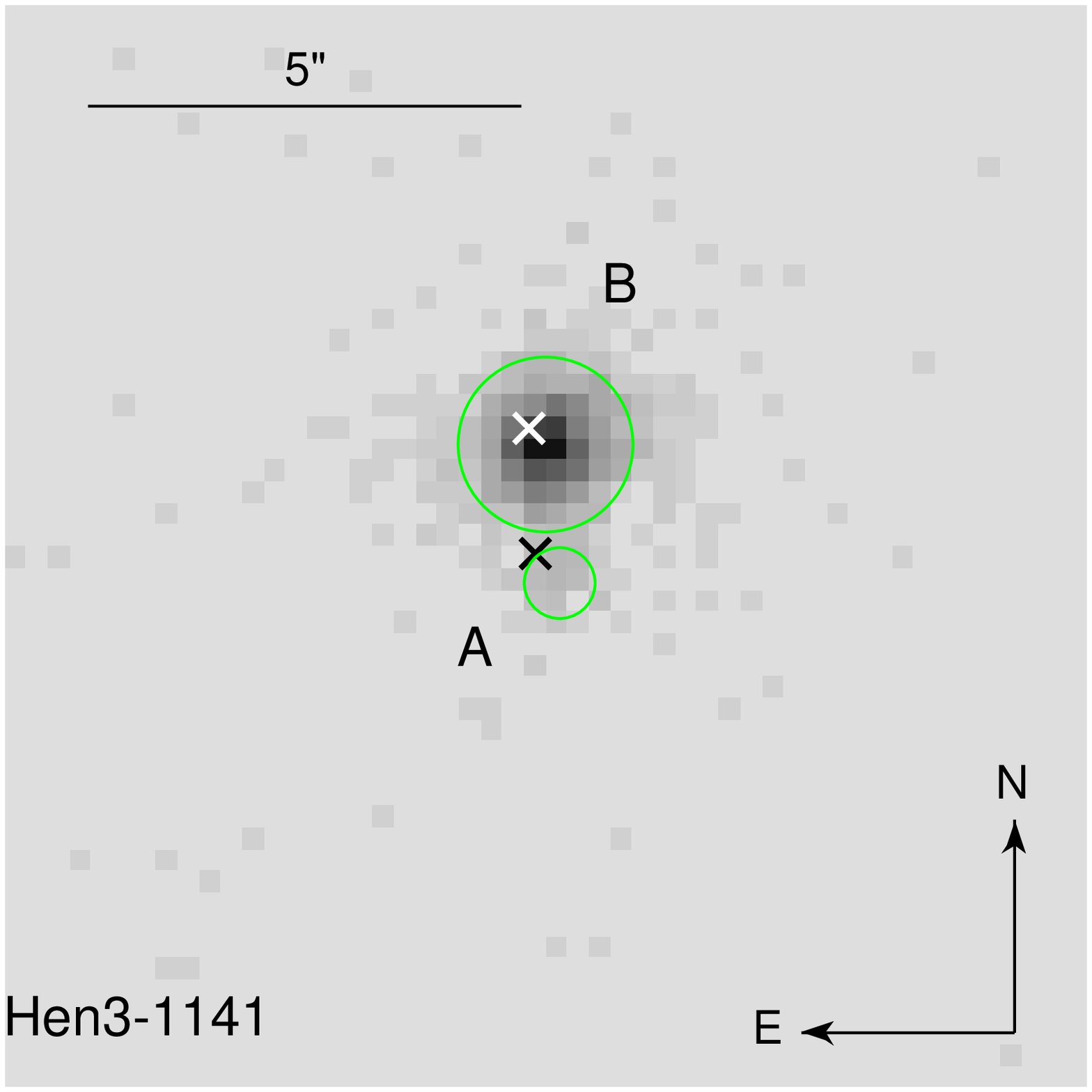}
       }
    }
  }
  \caption{{\em left} - X-ray flare energy number distributions for two samples of pre-MS stars in Taurus and Orion with bestfit power-laws \cite{Stelzer07.1}; {\em middle and right} - {\em Chandra} images of two Herbig stars (labeled 'A') and their known late-type companions ('B') [Stelzer et al. 2008, A\&A subm.]}
  \label{fig:2}
\end{figure}

Fig.~\ref{fig:2} shows the cumulative distribution of flare energies for two samples of 
pre-main sequence (pre-MS) stars in Orion and in Taurus. 
The power-law index is evaluated from the high-energy part of the distribution ($E > E_{\rm cutoff}$), 
which is not affected by incompleteness due to the sensitivity limit. 
This study came up with $\alpha \sim 2$ for both samples, supporting the nano-flare heating hypothesis 
but with uncertainties that do not excluce a flatter distribution \cite{Stelzer07.1}. 
A number of biases regarding the sensitivity to the detection of flares need to be taken into account: 
(i) short observations of variable stars are likely not to recover the true base level, and consequently 
the sensitivity for detecting flares is reduced; 
(ii) the ability to detect flares differs for stars of different mass, because higher-mass stars are 
brighter in X-rays, and brighter (i.e. more active) stars have more intense flares. 
A comparison of stars in Orion based on COUP and in the Cyg\,OB2 massive star forming
region taking into account the above caveats resulted in very similar flare frequency distributions
suggesting that these distributions are a universal phenomenon \cite{Albacete07.2}. 
A further caveat to be added is that, when compared to the Sun, the observed stellar flares have very 
large energies, and infering the nano-flare frequency from the above results depends on the validity of 
an extrapolation to small events.

\section{\centerline{The nature of coronae} \newline \centerline{in limiting regimes of stellar dynamos}}

The solar-type $\alpha\Omega$-dynamo operates in the interface between radiative core and convective envelope.  
Therefore, solar-like magnetic activity is believed to be limited to stars with interior structure
analogous to that of the Sun, and 
the solar-stellar analogy is expected to break down at the low- and high-mass extremes of the
stellar temperature sequence. This section discusses stars at those limits, i.e. fully radiative A/B-type stars
and fully convective very low-mass stars and brown dwarfs. Both groups comprise main-sequence objects
and young stars on the pre-MS.

\subsection{The fully radiative regime}

Despite no known mechanism for the production of X-rays exists in intermediate-mass stars, ever since the early
days of the {\em Einstein} Observatory a good fraction of them has been detected in X-rays. 
As the early X-ray missions had poor spatial resolution, it was argued that these stars might have 
late-type companions that are responsible for the X-ray emission but are not resolved from the A/B-type primaries 
\cite{Schmitt85.1}.

A recent archival study of {\em ROSAT} data of all A-type stars from the Bright Star Catalog,  
constituting the sample with the highest statistics every studied for this purpose, has shown that the 
X-ray detection fraction is $\sim 10-15$\,\% throughout all A/B spectral types \cite{Schroeder07.1}. 
No differences are found in the X-ray statistics of (magnetic) Ap stars with respect to normal A-type stars,
favoring the companion hypothesis. 

\cite{Czesla07.1, Schroeder08.1} have examined the possibility of a magnetic origin of the X-ray 
emission from A-type stars by comparing X-ray and magnetic field measurements. Most of the results support 
the companion hypothesis: 
(i) no difference between known doubles and presumed single stars in terms of X-ray luminosity vs. field strength,  
(ii) no difference in magnetic field strength for X-ray bright stars and upper limit sources, 
(iii) several arcseconds 
offset between the {\em ROSAT} X-ray source and the optical position of the A-type stars. 
On the other hand, in a high spatial resolution imaging study with {\em Chandra} 
many A- and B-type stars are X-ray sources although resolved from all known late-type companions \cite{Stelzer06.2}. 

Similar studies were lately carried out for intermediate-mass stars on the pre-MS 
(Herbig stars) with a surprising $100$\,\% detection rate for the primary in a sample of 
$9$ Herbig binaries or multiples [Stelzer et al. 2008, A\&A subm.]; see Fig.~\ref{fig:2}.   
Contrary to the case of the main-sequence (MS) 
B/A-type stars, there is a variety of possible scenarios for X-ray production
of Herbig stars: Next to the possibility of companions, these stars have relatively strong winds, 
some are accreting, and last but not least, magnetic fields have been detected on about $5-10$\,\% of Herbig 
stars. The nature and geometry of these fields is not yet known. Detailed X-ray diagnostics from high-resolution
spectroscopy, available so far for only one Herbig star \cite{Telleschi07.2},  
is needed to examine the source density, temperature and variability, parameters that are crucial to
distinguish between different emission mechanisms.

\subsection{The fully convective regime}

According to standard evolutionary models, MS stars with spectral type later than  $\sim$\,M3 are fully 
convective, and if any magnetic activity is to be maintained, 
the $\alpha\Omega$-dynamo must be replaced by alternative mechanisms for field generation.
For {\em young} very-low mass stars and brown dwarfs contributions to the X-ray emission from accretion
may be suspected. Among the four brown dwarfs 
in the TW\,Hya association the only known accretor 2M\,1207 has a very
faint upper limit to its X-ray luminosity, while the non-accretor TWA\,5B is a relatively bright X-ray source
(see Table~\ref{tab:twa}).  
Albeit not statistically sound, these observations are in line
with results for higher mass T\,Tauri stars (TTS), 
where weak-line TTS are on average X-ray brighter than classical TTS \cite[e.g.][]{Preibisch05.1}. 
\begin{table}
\begin{tabular}{lrrc} \hline
Designation & $W_{\rm H\alpha;10\,\%}$ & $L_{\rm x}$    & Refs \\ 
            & [km/s]                   & [erg/s]        & \\ \hline
TWA\,5B     & $162$                    & $4\,10^{27}$    &  \cite{Mohanty03.2, Tsuboi03.1} \\ 
2M\,1207    & $170...320$              & $<1.2\,10^{26}$ &  \cite{Stelzer07.2, Gizis04.1} \\
2M\,1139    & $111$                    & $...$           &  \cite{Mohanty03.2} \\
SSSPM\,1102 & $194$                    & $<5.3\,10^{26}$ &  \cite{Mohanty03.2,Stelzer07.2} \\
\hline
\end{tabular}
\caption{$10$\,\% width of H$\alpha$ emission as accretion proxy and X-ray luminosity as activity for all known brown dwarfs in TWA.}
\label{tab:twa}
\end{table}

\begin{figure}
  \parbox{\textwidth}{
  \parbox{0.5\textwidth}{
  \includegraphics[width=0.45\textwidth,angle=90]{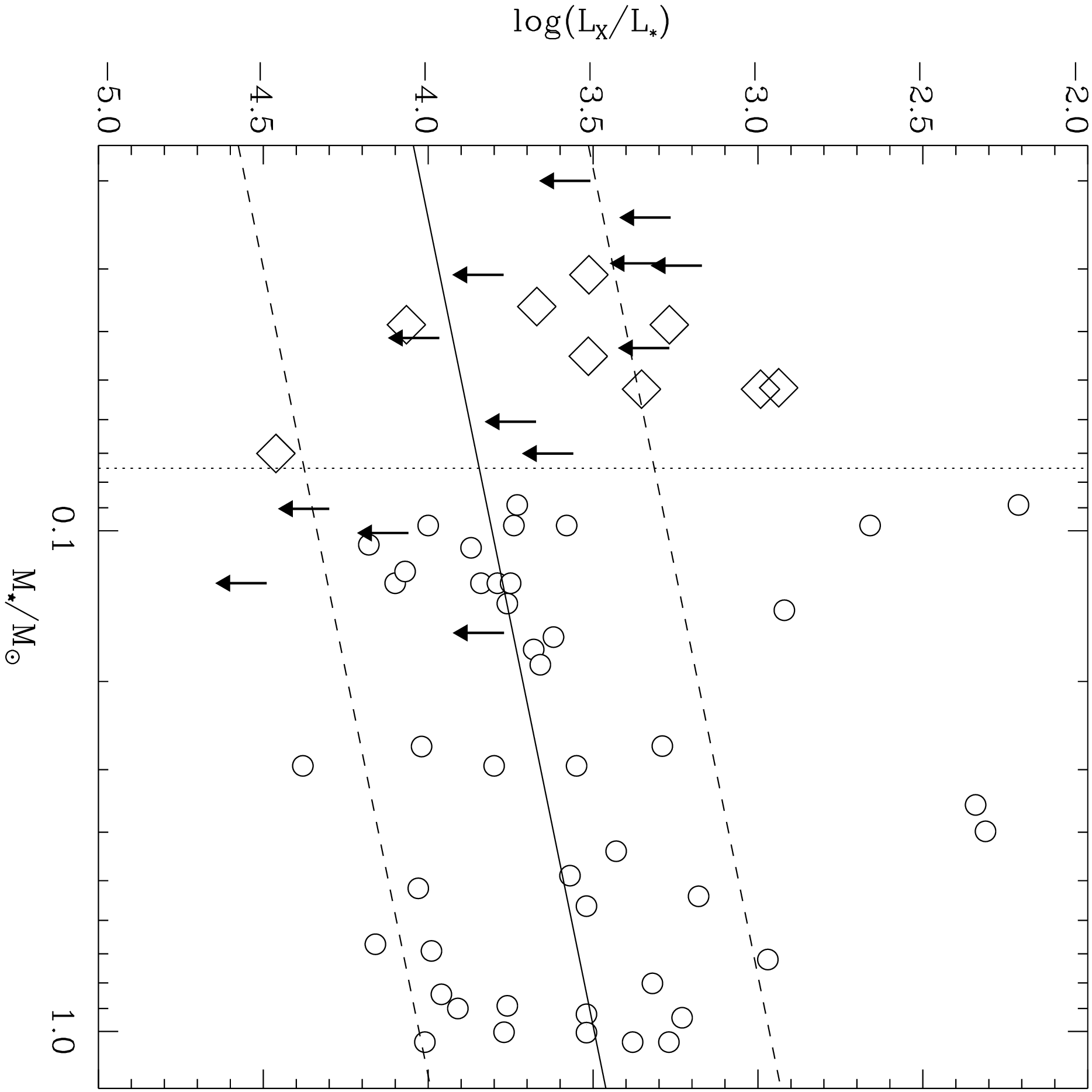}
  }
  \parbox{0.5\textwidth}{
  \includegraphics[width=0.5\textwidth]{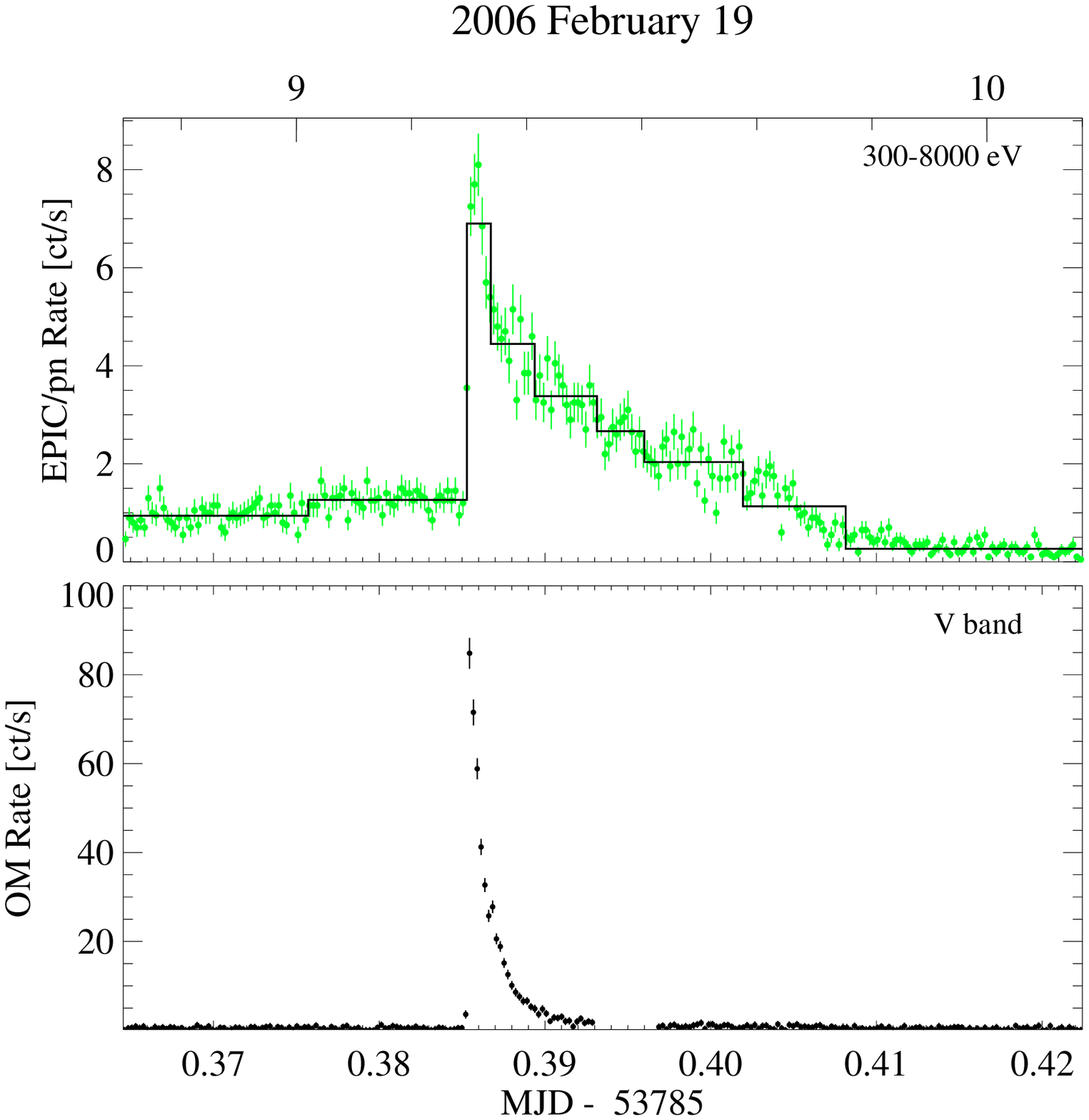}
  }
  }
  \caption{{\em left} - $L_{\rm x}/L_{\rm bol}$ vs. mass for low-mass stars and brown dwarfs in the Taurus 
Molecular Clouds \protect\cite{Grosso07.1}, {\em right} - giant flare on the M8 dwarf LP\,412-31 observed
simultaneously in X-rays and in the $V$ band with {\em XMM-Newton} \cite{Stelzer06.4}.}
  \label{fig:3}
\end{figure}

Larger samples of young brown dwarfs in the same star forming environment were observed e.g. during COUP and XEST,
allowing to compare their fractional X-ray luminosities $L_{\rm x}/L_{\rm bol}$ to those of higher-mass TTS. 
When the least-square fits of $L_{\rm x}/L_{\rm bol}$ vs. mass obtained for the objects above the 
substellar limit is extrapolated into the brown dwarf regime it is in reasonable agreement with the data
 (Fig.~\ref{fig:3}), suggesting that there is
no dramatic change in the activity level of brown dwarfs with respect to low-mass stars at ages of few Myrs. 
This can possibly be explained by the fact that both types of objects are fully convective, such that  
presumably the same kind of dynamo is at work. Better constraints on the large fraction of upper limits among
the brown dwarfs are needed to confirm these results. 

Very few {\em evolved} ultracool dwarfs, i.e. stars and brown dwarfs with spectral type later than M7, 
have been detected
in X-rays so far. Most of these detections are ascribed to flare events. 
A remarkable flare observation of an ultracool dwarf was obtained with {\em XMM-Newton} \cite{Stelzer06.4}. 
Thanks to the Optical Monitor onboard the X-ray satellite a giant flare on the M8 dwarf LP\,412-31 
was observed simultaneously in X-rays and in the $V$ band ($6$ mag brightening); see Fig.~\ref{fig:3}. 
The optical peak preceeded the maximum of the X-ray lightcurve by a few tens of seconds, 
in agreement with expectations from the standard flare scenario where chromospheric heating (giving rise to 
the optical flare) preceeds the filling of the coronal (X-ray bright) loops. 
Time-resolved spectroscopy during the decay of this event allowed to study 
the time evolution of temperature and emission measure. Comparison  
to hydrodynamic flare models yields a very long loop ($\sim 1\,R_\star$). 
Given the short flare rise time and the large loop size, the upward movement of the plasma likely involved 
high velocities that should give rise to line shifts observable with future X-ray missions.

\section{\centerline{Hot topics} \newline \centerline{on X-ray emission from pre-main sequence stars}}

\subsection{High-resolution spectroscopic observations}

{\em XMM-Newton} and {\em Chandra} have revolutionized our picture of stellar X-ray emission
thanks to their high-resolution X-ray spectrometers that offer novel capabilities for X-ray diagnostics
based on emission line analysis. 
Among the most compelling findings concerning cool stars was 
the identification of a high-density ($\sim 10^{12}\,{\rm cm^{-3}}$) and soft
excess (few MK) plasma in some classical TTS from an analysis of He-like triplets and 
O\,VII\,r / O\,VIII\,Ly$\alpha$ flux ratios, respectively: 
\begin{itemize}
\item The {\em XMM-Newton}/RGS spectrum of the prototypical high-density source TW\,Hya 
\cite{Stelzer04.3} 
was successfully modeled with a composite of a normal stellar corona and 
a one-dimensional steady state representation of an accretion shock \cite{Guenther07.2}. 
Anomalously low $f/i$ ratios have confirmed high densities in nearly all classical TTS observed at high spectral
resolution, the exception being T\,Tau (see below). 
\item Bright oxygen lines are useful temperature diagnostics for stellar X-ray sources. 
Comparison of the absorption corrected line luminosities revealed an excess
of flux in the O\,VII resonance line for classical TTS with respect to the expected value from 
the O\,VII\,r / O\,VIII\,Ly$\alpha$ flux ratios measured for 
weak-line TTS and MS stars \cite{Guedel07.4}. 
\end{itemize}

To summarize, in addition to a "normal" corona 
the X-ray emitting plasmas of classical TTS have a low-temperature component along with 
high densities, both characteristic for the physical conditions in TTS accretion shocks. 
In the case of T\,Tau there is a soft excess but no evidence for high density \cite{Guedel07.2} 
giving rise to speculations about a suppression of coronal heating by the accretion process. Alternatively, 
its X-ray emission might have a different origin, such as a stellar jet. 
X-ray emission was, indeed, imaged with {\em Chandra} from the pre-MS jet of DG\,Tau \cite{Guedel08.1}.
These observations are in qualitative agreement with 
hydrodynamic simulations that 
predict a localized X-ray source from a shock at several AU from the jet launching site \cite{Bonito07.1}.

\subsection{Effects of YSO X-rays on circumstellar matter}

X-ray emission from Young Steller Objects (YSOs) 
may be crucial for the structure and evolution of the circumstellar environment: 
(i) YSO X-rays are considered a prime ionization agent for stellar accretion disks
and are held responsible for Fe\,K$\alpha$ and [Ne\,II] emission features that have lately
been observed in a small number of objects; 
(ii) X-ray irradiation of close-in planets increases their mass loss rates and may lead
even to complete erosion of their atmospheres.

\subsubsection{Fe K$\alpha$ emission}

\begin{figure}
  \parbox{\textwidth}{
  \parbox{0.5\textwidth}{
  \includegraphics[width=0.4\textwidth,angle=270]{./giardino07_f5.ps}
  }
  \parbox{0.5\textwidth}{
  \includegraphics[width=0.5\textwidth]{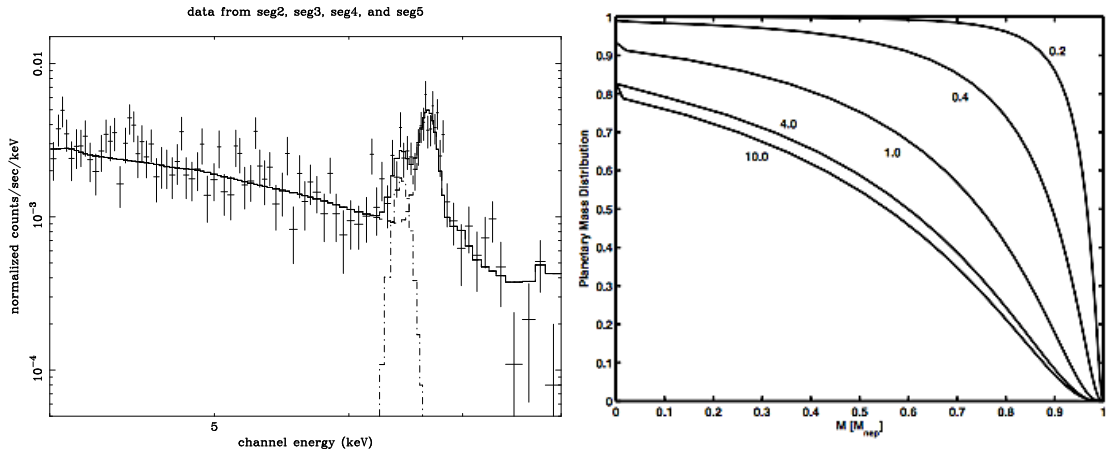}
  }
  }
  \caption{{\em left} - {\em XMM-Newton} EPIC spectrum with Fe\,K$\alpha$ line for the Class\,I protostar Elias\,29 observed during DROXO \protect\cite{Giardino07.1}; 
{\em right} - Calculated mass loss of a close-in ($0.02$\,AU) Neptune-mass exoplanet with $\rho = 2\,{\rm g/cm^3}$ around a dG star at different ages of the system in Gyr \protect\cite{Penz08.2}.}
  \label{fig:4}
\end{figure}

Fe K$\alpha$ emission has been observed for several decades during solar flares. 
The time-profile of the $6.4$\,keV line from the Sun was shown to be very similar to that
of soft thermal broad-band X-rays, and clearly distinct from the hard X-ray burst at the beginning
of the flare \cite{Culhane81.1}. This was taken as evidence that the observed Fe\,K$\alpha$ emission represents
fluorescence of cold material illuminated by the corona. A natural emission site is the stellar
photosphere. The same interpretation was applied to the Fe\,K$\alpha$ lines observed with
{\em Chandra} in (no more than) two evolved stars \cite{Osten07.1, Testa08.1}. 
In contrast, Fe~K$\alpha$ emission from pre-MS stars is 
believed to be emitted from the circumstellar disk as response to X-ray irradiation from
the central star \cite{Tsujimoto05.1}. The interpretation as fluorescence emission has been
questioned by \cite{Giardino07.1} who found that the presence of the Fe\,K$\alpha$ line 
of the Class\,I source Elias\,29 (shown in Fig.~\ref{fig:4}) is not related to the X-ray luminosity of the star.

\subsubsection{[Ne\,II]$12.8\,\mu$m emission}

The K-shell ionization energy of Ne coincides with the peak of typical stellar
X-ray spectra ($\sim 0.9$\,keV). Therefore, coronal X-rays have a high efficiency for 
photo-ionizing the Ne in stellar disks. 
Furthermore, X-ray irradiation produces a warm atmosphere
above circumstellar disks with temperatures of several $1000$\,K, comparable to the characteristic 
temperatures of the infrared fine-structure transitions of low-ionized Ne \cite{Meijerink08.1}. 
The calculations by \cite{Meijerink08.1} predict a direct correlation between [Ne\,II] line flux 
and X-ray luminosity. 
First observational studies searching for this correlation in limited samples 
yielded controversial results \cite{Pascucci07.1, Lahuis07.1}. A systematic study based on
the {\em Deep Rho Ophiuchus XMM-Newton Observation} (DROXO) suggests a rather complex situation
with probable influence of (circum)stellar parameters such as stellar mass and accretion rate on the
Ne emission; see Flaccomio et al. in this proceedings.

\subsubsection{Photoevaporation of planetary atmospheres}

The evaporation rate of a planet atmosphere depends on the temperature in its exosphere, 
which may be increased by UV/X-ray irradiation from the planet host star. 
The irradiating high-energy flux
as a function of time is therefore a crucial input to calculations of the evolution of 
atmospheres of close-in exoplanets. 
Scaling laws for the stellar X-ray luminosity as a function of age can be extracted from 
"The Sun in time" program \cite{Ribas05.1},
consisting of about a dozen solar-analogs, i.e. early G-stars from ages of
100Myr to 6.7Gyr, or from the X-ray luminosity functions of open clusters with different ages. 
The latter approach allows to take account of the spread in $L_{\rm x}$ of stars at a given age. 

The atmospheric mass loss rate
depends on the planet mass, its density, and its orbital separation  
(that determines the received X-ray flux). 
\cite{Penz08.1, Penz08.2} calculated the planet mass distribution as a function of these parameters and 
the (irradiation) time. 
An example of the results is shown in Fig.~\ref{fig:4}.
Evidently, most of the mass loss takes place during the first Gyr when the X-ray luminosity is highest. 
To summarize, 
(i) the mass loss decreases with increasing size of the orbit, and becomes negligible for Neptune-mass 
planets beyond about $0.1$\,AU, 
(ii) at a given separation the mass loss effects are much less pronounced for planets around
dM stars than for those with dG star hosts,
(iii) at a given separation the evaporation is stronger for low-density planets than for high-density planets, 
At a separation of $0.02$\,AU about the majority of Neptune-like planets with low atmospheric density 
get eroded to Super-Earths. 
  
All in all it seems that X-ray induced atmospheric evaporation could well be
responsible for the paucity of observed close-in high-mass planets. 
However, a comparison of these calculations to observations has yet to be done. 
Such a study must be based on unbiased samples with stars of various activity levels and 
planets with a wide range of mass and separations.





\bibliographystyle{aipproc}   

\bibliography{ms}

\IfFileExists{\jobname.bbl}{}
 {\typeout{}
  \typeout{******************************************}
  \typeout{** Please run "bibtex \jobname" to optain}
  \typeout{** the bibliography and then re-run LaTeX}
  \typeout{** twice to fix the references!}
  \typeout{******************************************}
  \typeout{}
 }

\end{document}